\documentclass[10pt,twoside,twocolumn,nofootinbib]{revtex4}

\usepackage{amsthm,amsmath,latexsym,amssymb,verbatim,enumerate,graphicx}
\usepackage{hyperref}
\usepackage{color}
\usepackage{anysize}
\usepackage{fancyhdr}
\usepackage{hyperref}
\usepackage{color}
\usepackage{tikz}
\usepackage{subeqnarray}
\usepackage{times,txfonts}
\usepackage{nicefrac}
\usepackage{ragged2e}
\usepackage{enumitem}

\DeclareMathAlphabet{\mathcal}{OMS}{cmsy}{m}{n} 

\marginsize{2cm}{2cm}{2cm}{2cm} 

\definecolor{darkblue}{rgb}{0.0, 0.0, 0.55}
\definecolor{darkgreen}{rgb}{0.0, 0.3, 0.0}
\definecolor{darkpurple}{rgb}{0.3, 0.0, 0.25}
\definecolor{darkorange}{rgb}{1.0, 0.549, 0.0}
\definecolor{darkred}{rgb}{0.5, 0.0, 0.0}
\definecolor{verylightgray}{rgb}{0.9, 0.9, 0.9}
\definecolor{gray75}{gray}{0.3}

\hypersetup{colorlinks=true, citecolor=darkred, linkcolor=darkblue, urlcolor=darkblue}

\definecolor{orcidlogocol}{HTML}{A6CE39}
\tikzset{
  orcidlogo/.pic={
    \fill[orcidlogocol] svg{M256,128c0,70.7-57.3,128-128,128C57.3,256,0,198.7,0,128C0,57.3,57.3,0,128,0C198.7,0,256,57.3,256,128z};
    \fill[white] svg{M86.3,186.2H70.9V79.1h15.4v48.4V186.2z}
                 svg{M108.9,79.1h41.6c39.6,0,57,28.3,57,53.6c0,27.5-21.5,53.6-56.8,53.6h-41.8V79.1z M124.3,172.4h24.5c34.9,0,42.9-26.5,42.9-39.7c0-21.5-13.7-39.7-43.7-39.7h-23.7V172.4z}
                 svg{M88.7,56.8c0,5.5-4.5,10.1-10.1,10.1c-5.6,0-10.1-4.6-10.1-10.1c0-5.6,4.5-10.1,10.1-10.1C84.2,46.7,88.7,51.3,88.7,56.8z};
  }
}

\newcommand\orcidicon[1]{\href{https://orcid.org/#1}{\mbox{\scalerel*{
\begin{tikzpicture}[yscale=-1,transform shape]
\pic{orcidlogo};
\end{tikzpicture}
}{|}}}}

\newcommand{\dd}{\mathrm{d}}		                     									
\newcommand{\var}[1]{\textrm{Var}{#1}}								

\newcommand{\cov}{\mathrm{Cov}}
\newcommand{\ii}{\mathrm{i}}
\newcommand{\ee}{\mathrm{e}}
\newcommand{\TT}{\mathrm{T}}
\newcommand{\supp}{\mathrm{Supp}\hspace{0.05cm}}
\newcommand{\diver}{\mathrm{div}\hspace{0.05cm}}

\theoremstyle{plain}																		

\newtheorem{thm}{Theorem}

\theoremstyle{definition}																	

\theoremstyle{remark}																		

\newtheorem*{rem}{Remark}

\begin{document}

\title[Nonextensive It\^{o}-Langevin Dynamics]{Nonextensive It\^{o}-Langevin Dynamics}

\author{Leonardo Santos\hspace{0.05cm}\href{https://orcid.org/0000-0002-8227-144X}{\includegraphics[scale=0.009]{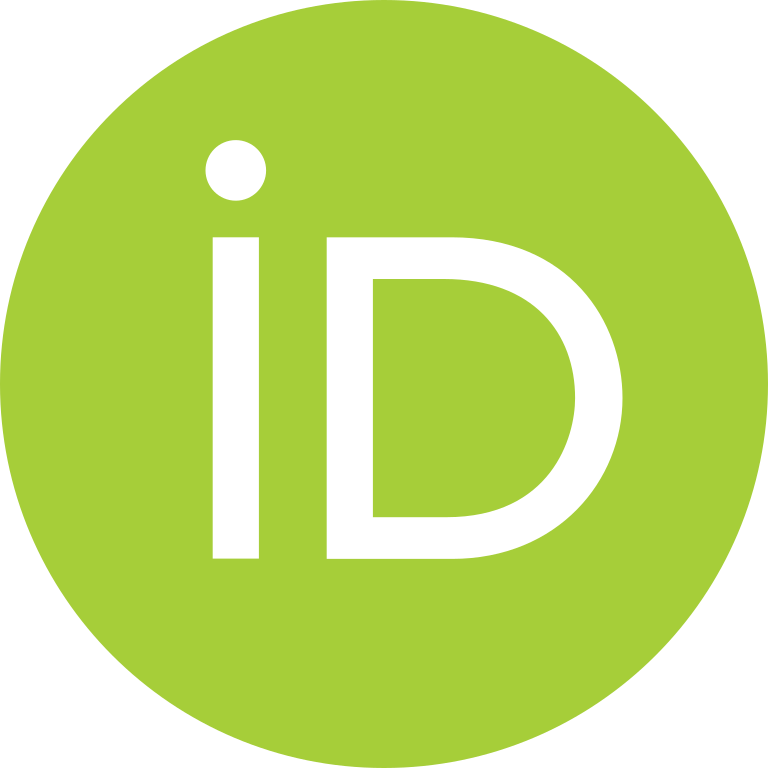}} }
\email{Leonardo.SSantos@uni-siegen.de}
\affiliation{Naturwissenschaftlich-Technische Fakultät, Universität Siegen, Walter-Flex-Straße 3, 57068 Siegen, Germany}
\affiliation{Departament of Mathematical Physics, Institute of Physics, University of São Paulo, São Paulo 05508-090, São Paulo, Brazil}

\date{\today}

\begin{abstract}
\noindent We study generalizations of It\^{o}-Langevin dynamics consistent within nonextensive thermostatistics. The corresponding stochastic differential equations are shown to be connected with a wide class of nonlinear Fokker-Planck equations describing correlated anomalous diffusion in fractals. A generalized central limit theorem is proposed in order to demonstrate how such equations emerge as a limit of correlated random variables. In doing so, we connect microscopic and macroscopic descriptions of correlated anomalous diffusion in a mathematically sound way and shed some light in explaining why $q$-Gaussian distributions appear quite often in nature. \\

\noindent Keywords: Nonextensive Statistical Mechanics, Tsallis Distributions, Nonlinear Fokker-Planck Equation, Langevin Dynamics, Central Limit Theorems.
\end{abstract}

\maketitle


\section{Introduction}

\noindent Since the Langevin's seminal contribution to the Brownian motion theory \cite{Langevin08}, stochastic differential equations (SDEs) have played a fundamental role in non-equilibrium thermodynamics \cite{UO30,Kubo66,GM13,TO15}. The Langevin's fundamental idea was to describe the Brownian particle motion in terms of the combination of deterministic forces, coming from damping and external potentials, and stochastic fluctuations, modeling collisions with fluid molecules. In a more general perspective, this work introduced a simple but powerful idea: the description of systems subjected to irregular and erratic fluctuations (noises) may be much more enlightening when modeled by an effective stochastic dynamics rather than Hamiltonian dynamics of many-body systems. Nowadays, Langevin's equations -- and their version in the It\^{o} stochastic calculus formalism, the It\^{o}-Langevin equations (ILEs) -- are workhorses of stochastic modeling in various branches of science, both in classical and quantum domains \cite{CK96,Gardiner,GZ99,BP12}.

The ILEs can be applied to a variety of problems, however their simplifying power is manifested mainly in the description of memoryless (Markovian) stochastic dynamics. The paradigmatic example is the diffusion of colloidal particles under the action of a memoryless noise. From a macroscopic point of view, the application of Fick's law together with local conservation of particles yields the well-known linear Fokker-Planck equation (FPE) \cite{Risken}. This system can be equivalently described by a local-in-time SDE, the \textit{standard} It\^{o}-Langevin equation
\begin{eqnarray}\label{eq:ito_langevin}
\dd X_t=F(X_t,t)\dd t+\sqrt{2D}\dd W_t,
\end{eqnarray}
where $X_t$ is a stochastic process representing the system state, $F$ is the deterministic force acting on the system, $D$ is the diffusion matrix and $W_t$ is the \textit{standard} Wiener process. These equations typically yields a Gaussian distributions. In particular, in the free diffusion regime, the covariance matrix linearly depends on time. These two properties define the Gaussian normal diffusion regime, a signature of memoryless dynamics.

The standard ILE (\ref{eq:ito_langevin}) provides a satisfactory description for the stochastic dynamics of a wide variety of systems \cite{Kampen81}. This is because we often deal with situations where the noises come from weak and uncontrollable interactions with the surrounding environment. In general, such environments are much larger, complex and chaotic than the analyzed system. This prevents system information that flowed into the environment from returning at a later time, in such a way that the system dynamics is (at least approximately) Markovian. Similarly, invoking the central limit theorem (CLT) \cite{Durrett10}, we would expect a Gaussian process modelling the effective interaction. Significant deviations from this regime often occur in complex systems \cite{Vemuri78}, which implies that the stochastic dynamics may be neither Markovian nor Gaussian. In this case, the ILE needs to be generalized, e.g. by introducing memory kernels and/or multiplicative noises \cite{dosSantos19}.

In this paper, we propose a generalization of the standard It\^{o}-Langevin dynamics consistent within nonextensive (NE) statistical mechanics -- a possible generalization of the Boltzmann-Gibbs (BG) theory based on Tsallis non-additive entropy \cite{Tsallis88}. The motivation for such a generalization is manifold. Plastino-Plastino \cite{PP94} demonstrated that equilibrium distributions for systems in contact with a \textit{finite} thermal bath obey the Tsallis statistics rather than BG (which emerges only in the limit of an infinite thermal bath). It was demonstrated by Plastino-Plastino \cite{PP95} and Tsallis-Bukman \cite{TB96} that normalized scale invariant solutions of a power-law-type nonlinear FPE are distributions that maximize Tsallis entropy. Such distributions also emerge in the weak chaos regime \cite{K12,TB16,RTBT17a,RTBT17b,CTB22}, in a wide class of systems featuring long-ranged interactions \cite{AT98,CAT14,CTB14}, in anomalous transport in optical lattices \cite{Lutz03,LR13}, anomalous diffusion in granular media \cite{CRSA15}, diffusion of Hydra cell in cellular aggregates \cite{URGS01}, finances \cite{Borland02}, high-energy experiments \cite{CMS10a,CMS10b,ALICE10}, etc. \cite{TsallisTemuco}. 

Based on the mathematical formalism of $q$-statistics \cite{UT22}, we propose a wide class of SDEs that generalize the ILEs (\ref{eq:ito_langevin}). The formalism is based on generalizations of the standard Wiener process and It\^{o} stochastic calculus, in such a way that the corresponding SDEs are local-in-time and can be solved and/or simulated in a similar fashion as Eq. (\ref{eq:ito_langevin}). As particular cases, these equations describe the stochastic dynamics of linear and a wide class of nonlinear FPEs describing correlated anomalous diffusion in fractals, without multiplicative noises nor couplings with the macroscopic evolution, extending ideas presented in Ref. \cite{Santos21} and in contrast to the approaches of Refs. \cite{Borland98,Frank01}. Furthermore, we demonstrate how such processes can emerge as a limit of correlated random variables by extending the generalized CLT by Umarov-Tsallis-Steinberg (UTS) \cite{UTS08}. In addition to supporting NE stochastic dynamics, we believe that this last result also sheds light on explaining why $q$-Gaussians appear so frequently in nature.

We have organized the paper as follows. In Sec. \ref{sec:2} we present the mathematical foundations of our formalism. We have chosen to present such results in a concise manner where the technical details are discussed in Appendix \ref{app:A}. In Sec. \ref{sec:3} we discuss our generalization of the It\^{o}-Langevin dynamics. We finish the paper with a short discussion in Sec. \ref{sec:4}. 

\section{Mathematical Foundations}\label{sec:2}

\subsection{$\theta$-Gaussian Distributions}

\noindent The Tsallis entropy $S_q$ of a real random variable $X$, taking values in a $d$-dimensional space with probability density function (PDF) $p$, is defined through
\begin{eqnarray}
S_q[X]=\int_{\supp p}p(x)\log_q\frac{1}{p(x)}\dd x,
\end{eqnarray}
where $q$ is a real parameter and $\log_q u=(u^{q-1}-1)/(q-1)$, $u\geq 0$, is the $q$-logarithm. The Boltzmann-Gibbs (BG) entropy is recovered when $q\to 1$. The maximization of $S_q$, under appropriated norm and width constrains, yields a generalized Gaussian distribution often referred to as $q$-Gaussian distribution \cite{CT91,Tsallis09,GT04} 
\begin{eqnarray}\label{eq:theta_Gaussian}
G_q(x;\mu,\Sigma)=\frac{\exp_q\left[-\frac{1}{2_{qd}}(x-\mu)^\TT \Sigma^{-1}(x-\mu)\right]}{\sqrt{(2_{qd}\pi_{qd})^d\det \Sigma}},
\end{eqnarray}
where $\exp_q u=[\max\{0,1+(1-q)u\}]^{\frac{1}{1-q}}$ is the $q$-exponential, and $2_{qd}$ and $\pi_{qd}$ are normalization parameters such that $2_{qd}\to 2$ and $\pi_{qd}\to \pi$ when $q\to 1$, thus recovering the usual Gaussian distribution in such a limit. In this paper, we consider a generalization of $G_q$ defined through
\begin{eqnarray}\label{eq:q_Gaussian}
G_{\theta}(x;\mu,\Sigma)=\frac{\exp_q\left\{-\frac{1}{2_{\theta}}[(x-\mu)^\TT\Sigma^{-1}(x-\mu)]^\eta\right\}}{\sqrt{(2_\theta\hspace{0.05cm} \pi_\theta)^d\det\Sigma}},
\end{eqnarray}
where $\theta=(q,\eta,d)\in \mathbb{R}^2\times \mathbb{N}$. Clearly, $G_{\theta}\to G_q$ when $\eta\to 1$ if we set $2_{\theta}\to 2_{qd}$ and $\pi_{\theta}\to \pi_{qd}$ in such a limit. We denote $X\sim N_\theta(\mu;\Sigma)$ to refer to a $\theta$-Gaussian random variable with expectation value $\mu=E[X]$ and covariance matrix $\Sigma=E[XX^\TT]$. We shall only consider those values of $\theta$ such that $G_\theta$ is normalized and $\mathrm{Tr}\hspace{0.05cm} \Sigma<\infty$. A short calculation shows that such conditions are fulfilled if, and only if, $\eta>0$ and $(d+2)(q-1)<2\eta$. 

\begin{figure*}
    \centering
    \includegraphics[scale=0.3825]{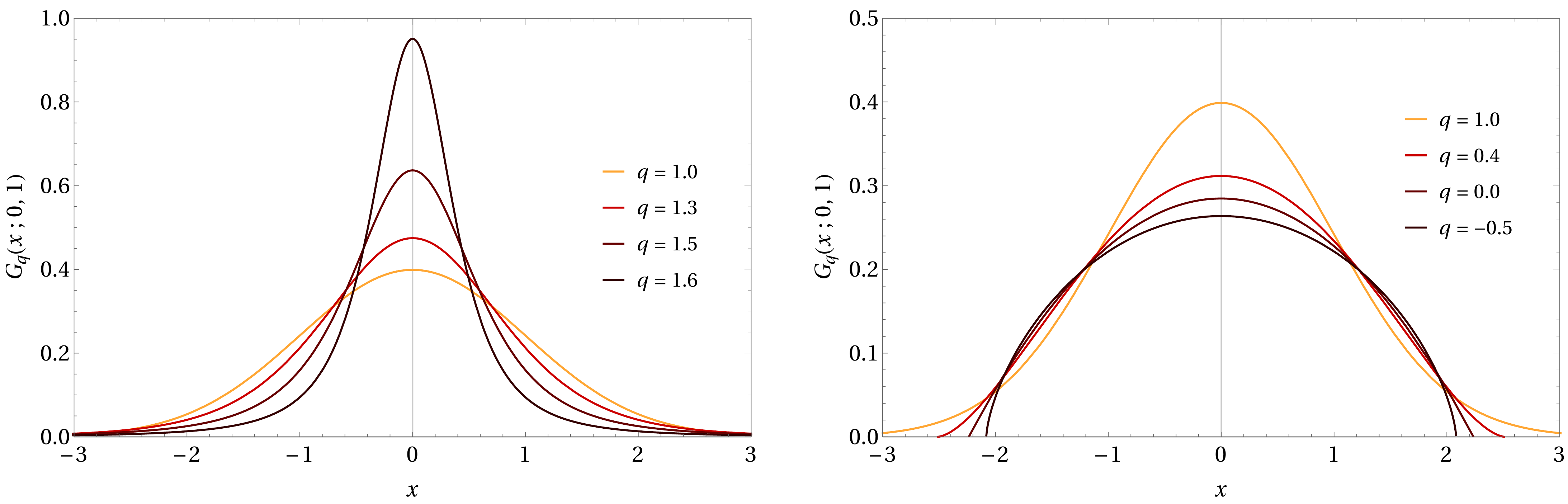}
    \caption{One dimensional $q$-Gaussians with $\mu=0$ and $\sigma^2=1$ [Eqs. (\ref{eq:q_Gaussian}) and (\ref{eq:theta_Gaussian_1d})].}
    \label{fig:1d_qgauss}
\end{figure*}

Both parameters $q$ and $\eta$ modify the properties of $G_\theta$ in relation to the Gaussian distribution. The parameter $q$ is more significant in this sense as it breaks the exponential decay when $q>1$ and limits the support of the distribution when $q<1$. Indeed, for $q\geq 1$, the support of $G_\theta$ is the whole $\mathbb{R}^d$ whereas for $q<1$ it has a compact support given by $\supp G_\theta=\{(x-\mu)^\TT \Sigma^{-1}(x-\mu)<(1-q)^{-1/\eta}\}$. For the one-dimensional case, Eq. (\ref{eq:theta_Gaussian}) reduces to
\begin{eqnarray}\label{eq:theta_Gaussian_1d}
G_\theta(x;\mu,\sigma^2)=\frac{1}{\sqrt{2_\theta\hspace{0.05cm}\pi_\theta\hspace{0.05cm} \sigma^2}}\exp_q\left(-\frac{|x-\mu|^{2\eta}}{2_\theta\hspace{0.05cm}\sigma^2}\right),
\end{eqnarray}
where $\sigma^2$ denotes variance. This distribution has a power-law asymptotic behavior $G_\theta\sim x^{\frac{2\eta}{1-q}}$ if $q>1$ and its support is the interval defined by $|x-\mu|^\eta<(1-q)\sigma^2$ if $q<1$ (see Fig. \ref{fig:1d_qgauss}).

\subsection{Central Limit Theorems}

\noindent As already mentioned, $\theta$-Gaussian distributions appear in a variety of situations in complex systems \cite{PMMS09}. In the last decades, several models have been proposed to explain the ubiquity of such distributions in nature \cite{Tsallis01,Tsallis05,MTG06}. In this paper, we shall demonstrate here how these PDFs can emerge as the limit of correlated random variables from a generalization of the central limit theorem (CLT) based on the Umarov-Tsallis-Steinberg (UTS) approach \cite{UTS08} (see also Ref. \cite{UTGS10}). The first definition we introduce in this direction is that of a $\vartheta$-characteristic function of a random variable $X$, which is the $\vartheta$-Fourier transform (FT) of its PDF $p$: 
\begin{eqnarray}
\textrm{F}^{\hspace{0.025cm}\vartheta}_{X}(k)=\int_{\supp p} p(x)\exp_q\left\{\ii\hspace{0.025cm} (k^\TT x)\hspace{0.025cm}[p(x)]^{\frac{1}{d}(1-\alpha)}\right\}\dd x,
\end{eqnarray}
where $\vartheta=(\theta,\lambda)$, $\lambda\in\mathbb{R}$, and
\begin{eqnarray}
\alpha(\vartheta)=\eta+d\lambda(1-q),
\end{eqnarray}
$\alpha\equiv \alpha(\vartheta)$. This integral transform generalizes the usual FT and UTS' $q$-FT. In what follows two properties of $\textrm{F}^{\hspace{0.025cm}\vartheta}$ shall be relevant. First, the scaling property:
\begin{eqnarray}
\textrm{F}_{r X}^{\hspace{0.025cm}\vartheta}(k)=\textrm{F}_{X}^{\hspace{0.025cm}\vartheta}(r^{\alpha}k),
\end{eqnarray}
$r>0$. The other property is that the $\vartheta$-characteristic function of a $\theta$-Gaussian random variable $X\sim N_\theta(0,\Sigma)$ is of the form
\begin{eqnarray}
\textrm{F}^{\hspace{0.025cm}\vartheta}_{X}(k)=\mathrm{Exp}_{\vartheta}\left[-\frac{2_\theta}{4(2_\theta \pi_\theta)^{1-\alpha}}\frac{k^\TT \Sigma k}{(\det \Sigma)^{\frac{1}{d}(1-\alpha)}}\right],
\end{eqnarray}
where $\mathrm{Exp}_{\vartheta}(x)$ is a non-negative real function such that $\mathrm{Exp}_{\vartheta}(0)=1$ (see Appendix \ref{app:a1}).

It is often useful to describe power-law decaying PDFs in terms of escort distributions and moments \cite{BS93,TPA09}. We define the escort distribution of order $s$ of a random variable $X$ with PDF $p$ as
\begin{eqnarray}
p_s(x)=\frac{[p(x)]^s}{v_s[p]},
\end{eqnarray}
where
\begin{eqnarray}
v_s[p]=\int_{\supp p}[p(x)]^s\dd x.
\end{eqnarray}
Similarly, the $s$-expectation value is defined as the expectation value with respect to the escort distribution of order $s$. A short calculation shows that the $\vartheta$-FT of $X$ can be expanded in terms of its escort moments:
\begin{eqnarray}
\textrm{F}^{\hspace{0.025cm}\vartheta}_{X}(k)=1&+&\ii v_{s_1}[p] k^\TT E_{s_1}[X]\nonumber \\ &-&\frac{q\hspace{0.05cm}v_{s_2}[p]\hspace{0.05cm}k^\TT}{2} E_{s_2}[XX^\TT]k+\dots,
\end{eqnarray}
where $s_n=1+n(1-\alpha)/d$, $n\geq 1$. 

We now introduce a notion of independence for correlated random variables, the $\vartheta$-independence. Given a sequence of random variables $X_n$, taking values in some $\mathbb{R}^d$ with sufficiently well-behaved PDFs, this sequence is $\vartheta$-independent if for all $n\geq 1$ the $\vartheta$-characteristic function of the sum $X_1+\dots +X_n$ $\vartheta$-factorizes as follows:
\begin{eqnarray}
\textrm{F}^{\hspace{0.025cm}\vartheta}_{X_1+\dots+X_n}(k)=\textrm{F}^{\hspace{0.025cm}\vartheta}_{X_1}(k)\hspace{0.05cm}\otimes_{\vartheta}\hspace{0.05cm} \dots \otimes_{\vartheta}\hspace{0.05cm} \textrm{F}^{\hspace{0.025cm}\vartheta}_{X_n}(k),
\end{eqnarray}
where $\otimes_{\vartheta}$ is defined as
\begin{eqnarray}
a\otimes_{\vartheta} b=\mathrm{Exp}_{\vartheta}\left(\mathrm{Log}_{\vartheta}\hspace{0.025cm}a+\mathrm{Log}_{\vartheta}\hspace{0.025cm}b\right),
\end{eqnarray}
$a,b>0$ and $\mathrm{Log}_{\vartheta}$ is the $\vartheta$-logarithm, the inverse of $\mathrm{Exp}_{\vartheta}$. Finally, if a sequence $X_n$ of random variables is such that the sequence of $\vartheta$-characteristic functions $\textrm{F}_{X_n}^{\hspace{0.025cm}\vartheta}$ pointwise converges to $\textrm{F}_{X}^{\hspace{0.025cm}\vartheta}$, for a random variable $X$, we then say that $X_n$ $\vartheta$-converges to $X$.

We now consider a sequence of random variables $X_n$ that are $\vartheta$-independent and identically distributed with PDF $p$. We assume that
\begin{eqnarray}
\int_{\supp p} \|{x}\|^2[p({x})]^{s_2}\dd {x}<\infty,
\end{eqnarray}
Then it is possible to demonstrate (see Appendix \ref{app:a1}) that the sequence $\tilde{X}_n$,
\begin{eqnarray}
\tilde{X}_n=(nz)^{-\frac{1}{2\alpha}}\left(S_n-n E_{s_1}[X_n]\right),
\end{eqnarray}
$S_n=X_1+\dots +X_n$, $\vartheta$-converges to a $\theta$-Gaussian random variable ${X}$ with zero expectation value and covariance matrix ${\Sigma}$ satisfying
\begin{align}
\frac{\Sigma}{(\det\Sigma)^{\frac{1-\alpha}{d}}}=\frac{2_{\theta'}v_{s_2}[p]}{2_{\theta}(2_{\theta}\pi_{\theta})^{\alpha-1}}\left(\frac{\pi_{\theta} s_2}{\pi_{\theta'}}\right)^{\frac{d}{2}}E_{s_2}[X_1X_1^\TT],
\end{align}
where $\theta'=(q',\eta,d)$, $q'=[s_2-(1-q)]/s_2$, $\mathbb{I}_d$ denotes the $d\times d$ identity matrix and $s_n=1+n(1-\alpha)/d$. In short, the attractor of $\vartheta$-independent random variables is the $\theta$-Gaussian distribution. The standard CLT is recovered when $q=\eta\to 1$. When $\eta=1$ but $q\neq 1$, we have a multivariate version of UTS $q$-CLT \cite{UTS08} (cf. Ref. \cite{UT07}). 

\subsection{Generalized Wiener Processes}

\noindent The generalized CLT discussed above has several possible consequences worthy to be explored. From now on, however, we shall focus only on generalizations of the Wiener processes. In particular, we define $Q_t^{\vartheta}$, for $1\leq q<\min\{(\lambda d+\eta)/\lambda d,(d+2\eta+2)/(d+2)\}$, as a stochastic process in $\mathbb{R}^d$ satisfying the following conditions:
\begin{enumerate}[label=(\arabic*)]
\item ${Q}_0^{\vartheta}={0}$ almost surely;
\item The paths $t\mapsto{Q}_{t}^{\vartheta}$ are continuous with probability one;
\item ${Q}_{t}^{\vartheta}-{Q}_{s}^{\vartheta}\sim N_\theta\big[0,(t-s)^{\frac{1}{\alpha}}\mathbb{I}_d\big]$, $0\leq s\leq t$;
\item ${Q}_{t}^{\vartheta}-{Q}_s^{\vartheta}$, $0\leq s\leq t$, is ${\vartheta}$-independent of ${Q}_s^{\vartheta}$;
\item Given $t,s\geq 0$, the covariance between the components of ${Q}_t^{\vartheta}$ and ${Q}_s^{\vartheta}$ satisfies \begin{eqnarray}\label{eq:cova_maintext}
2\hspace{0.05cm}\cov[Q_{t;\hspace{0.05cm}\mu}^{{\vartheta}},Q_{s;\hspace{0.05cm}\nu}^{{\vartheta}}]=\delta_{\mu\nu}(t^{\frac{1}{\alpha}}+s^{\frac{1}{\alpha}}-|t-s|^{\frac{1}{\alpha}}),
\end{eqnarray}
where $\delta_{\mu\nu}$ denotes the Kronecker's delta.
\end{enumerate}
In the Appendix \ref{app:a2} we demonstrate, for the cases where $\eta=1$, how such processes emerge as a limit of correlated random variables through the generalized CLT presented above.

It is easy to verify that the standard Wiener process ${W}_t$ corresponds to $q=\eta\to 1$. These $\vartheta$-generalizations differ from ${W}_t$ in fundamental aspects such as the fact that they have increments that are not Gaussian nor independent. However, some properties of ${W}_t$ are also present in ${Q}_t^\vartheta$, e.g. conditions \ref{item:i} and \ref{item:ii}. Also, the paths of ${Q}_t^{\vartheta}$ are in general continuous but not smooth and hence non-differentiable \cite{MP10}. In fact, notice that, given $\Delta t$ and $A$, both positive, the probability that $\|\Delta {Q}_t^{\vartheta}\|/\Delta t$ takes a value bigger than $A$  is
\begin{align}
\mathrm{Pr}\left[\left\|\frac{\Delta {Q}_t^\vartheta}{\Delta t}\right\|>A\right]&=\int_{\|{x}\|>A} G_\theta\big({x};{0},(\Delta t)^{\frac{1}{\alpha}-2}\mathbb{I}_d\big)\dd {x} \nonumber \\ &=\int_{\|{u}\|>(\Delta t)^{1-\frac{1}{2\alpha}}A} G_\theta({u};{0},\mathbb{I}_d)\dd {u}.
\end{align}
In the limit $\Delta t \to 0$, this probability equals one for any $A>0$ if $\vartheta=(q,\eta, d,\lambda)$ satisfies
\begin{eqnarray}
q<\frac{2\eta-1+2d\lambda}{2d\lambda}.
\end{eqnarray}
In other words, $\|\Delta{Q}^\vartheta\|/\Delta t$ is bigger than any $A>0$ with probability one and thus the paths of ${Q}_t^{\vartheta}$ cannot be differentiable. In addition, the paths of ${Q}_t^\vartheta$ obey the following self-similarity law:
\begin{eqnarray}\label{eq:self_similarity}
{Q}_{t}^\vartheta =c^{-\frac{1}{2\alpha}}{Q}_{t/c}^\vartheta,
\end{eqnarray}
for any $c>0$. The last two mentioned properties are also present in the standard Wiener process.

We finish this section by defining the stochastic integral of a function $f$ with respect to ${Q}_t^\vartheta$ as a stochastic process ${I}_t$,
\begin{eqnarray}
{I}_t=\int_0^t f\big({Q}_\tau^\vartheta,\tau\big)\dd {Q}_\tau^\vartheta,
\end{eqnarray}
whose PDF is obtained by
\begin{eqnarray}\label{eq:stochastic_integral}
\mathrm{F}^{\hspace{0.05cm}\vartheta}_{{I}_t}({k})=\lim_{n\to \infty}\bigotimes_{\vartheta; \hspace{0.05cm} [t_{j-1},t_j]\in\mathsf{p}_n} \mathrm{F}^{\hspace{0.05cm}\vartheta}_{f({Q}^{\vartheta}_{t_{j-1}},\hspace{0.05cm}t_{j-1})\Delta {Q}_{t_j}^\vartheta}({k}),
\end{eqnarray}
where $\mathsf{p}_n$ is a sequence of partitions of the interval $[0,t]$ with mesh going to zero, $\Delta {Q}^\vartheta_{t_j}\equiv {Q}^\vartheta_{t_j}-{Q}_{t_{j-1}}^\vartheta$, and the product in the right-hand-side is the $\vartheta$-product indexed by the elements of $\mathsf{p}_n$. This definition generalizes (in distribution) the usual stochastic integral (cf. Ref. \cite{Privault13}) with respect to the Wiener process. Moreover, it is possible to demonstrate (see appendix \ref{app:a3}) that this stochastic integral obeys a generalized Itô formula:
\begin{eqnarray}\label{eq:thetaito}
I_t\sim N_\theta\left[{0},\hspace{0.1cm}\left(\int_0^t |f(\tau)|^{2\alpha}\dd \tau\right)^{\frac{1}{\alpha}}\mathbb{I}_d\right],
\end{eqnarray}
where $f(t)\in L^{2\alpha}([0,t])$, $t\geq 0$.

\section{Nonextensive It\^{o}-Langevin Equations}\label{sec:3}

\noindent In this section we shall discuss a possible generalization of the standard It\^{o}-Langevin equation (\ref{eq:ito_langevin}) describing a stochastic system subjected to a non-Markovian $\theta$-Gaussian noise. The state of the system is represented by a stochastic process $X_t$ satisfying the following SDE:
\begin{eqnarray}\label{eq:NE_ito_langevin}
\dd X_t=F(X_t,t)\dd t+B(t)\dd W^{\vartheta}_t,
\end{eqnarray}
where $B$ is a (possible time dependent) $d\times d$ matrix, $W_t^{\vartheta}=\sqrt{\varepsilon_{\vartheta}}Q^{\vartheta}_t$, $\varepsilon_{\vartheta}$ is a non-negative number such that $\varepsilon_{\vartheta}\to 1$ when $q=\eta\to 1$, ensuring that the standard Wiener process $W_t$ and It\^{o}-Langevin equation (\ref{eq:ito_langevin}) are recovered in such a limit. The solutions of this equations can be extracted through the generalized It\^{o} formula (\ref{eq:thetaito}):
\begin{eqnarray}\label{eq:NE_ito_formula}
\int_0^t f(t')\dd W_t^ {\vartheta}\sim N_\theta\left[0,\hspace{0.05cm}\varepsilon_{\vartheta}\left(\int_0^t |f(\tau)|^{2\alpha}\dd \tau\right)^{\frac{1}{\alpha}}\right].
\end{eqnarray}

In general it is very difficult to solve Eq. (\ref{eq:NE_ito_langevin}) given a general deterministic force $F$, even in the standard form \cite{Gardiner}. In this paper, we shall focus mainly on linear systems, for which we write
\begin{eqnarray}\label{eq:linear_force}
F(x,t)=Ax+b(t),
\end{eqnarray}
where $A$ is $d\times d$ time independent matrix and $b(t)$ is a $d$-dimensional time dependent vector. We also make the reasonable assumption that the eigenvalues of $A$ have negative real part, thence guaranteeing the stability of the solutions. Plugging Eq. (\ref{eq:linear_force}) into Eq. (\ref{eq:NE_ito_langevin}), we can obtain (after some manipulations) the following expression 
\begin{eqnarray}
X_t=\ee^{At}X_0+\int_0^t \ee^{A(t-t')}b(t')\dd t+\int_0^t \ee^{A(t-\tau)}B(\tau)\dd W_{\tau}^{\hspace{0.025cm}\vartheta}.
\end{eqnarray}
From the application of Eq. (\ref{eq:NE_ito_formula}) in the above equation together with a point source initial condition centered at $\bar{x}_0$, i.e. $X_0=\bar{x}_0$ (almost surely), we conclude that $X_t$ is a $\theta$-Gaussian stochastic process with expectation value $\bar{x}(t)=E[X_t]$ and covariance matrix $\Sigma(t)=E[X_t X_t^\TT]$ satisfying
\begin{eqnarray}\label{eq:qexpectation_value_NE_Langevin}
\bar{x}(t)=\bar{x}_0+\int_0^t \ee^{A(t-\tau)}b(\tau)\dd \tau,
\end{eqnarray}
and
\begin{eqnarray}\label{eq:cov_matrix_NE_Langevin1}
\Sigma(t)=\varepsilon_\vartheta\left(\int_0^t \left\{\ee^{A(t-\tau)}[2D(\tau)]^{\frac{1}{\alpha}}\ee^{A^\TT(t-\tau)}\right\}^{\alpha}\dd \tau\right)^{\frac{1}{\alpha}},
\end{eqnarray}
where $D$, $2D=(BB^\TT)^{\alpha}$, is a positive semi-definite matrix, the diffusion matrix.

In the simplest possible situation we have the free diffusion scenario, where the system evolves in the absence of deterministic forces (i.e., $F=0$). Under this assumption, Eq. (\ref{eq:cov_matrix_NE_Langevin1}) reduces to 
\begin{eqnarray}\label{eq:cov_matrix_NE_Langevin2}
\Sigma(t)=\varepsilon_\theta\left(\int_0^t 2D(\tau) \dd \tau\right)^{\frac{1}{\alpha}}.
\end{eqnarray}
In the particular case where the diffusion matrix is time independent, the covariance matrix equals
\begin{eqnarray}\label{eq:Sigma_anomalous_diff}
\Sigma(t)=\varepsilon_\vartheta(2 Dt)^{\frac{1}{\alpha}}.
\end{eqnarray}
In summary, in the free diffusion scenario the stochastic system described by Eq. (\ref{eq:NE_ito_langevin}) corresponds to a $\theta$-Gaussian PDF with a power-law time-dependence in the covariance matrix. These two properties are signatures of a correlated anomalous diffusion regime, which is characterized by the exponent
\begin{eqnarray}\label{eq:anomalous_exp}
\alpha^{-1}=[\eta+d\lambda(1-q)]^{-1}.
\end{eqnarray}
Normal Gaussian diffusion corresponds to $q=\eta=1$, which yields $\alpha^{-1}=1$ in the above equation. Otherwise, the system is super-diffusive if $\alpha^{-1}>1$, sub-diffusive if $\alpha^{-1}<1$ and normal if $\alpha^{-1}=1$.

In the absence of a deterministic force, the system will never reach the equilibrium since the covariance matrix has a power-law dependence on time. If we consider a linear force $F=Ax+b(t)$, however, the system will eventually reach the equilibrium whenever $A$ is stable (i.e. its eigenvalues have negative real part) and $b$ is time independent. In these regime, the equilibrium expectation is read immediately from Eq. (\ref{eq:qexpectation_value_NE_Langevin}):
\begin{eqnarray}
\bar{x}_{\textrm{eq}}=\bar{x}_0+A^{-1}b.
\end{eqnarray}
To obtain the equilibrium covariance matrix, notice the time derivative of Eq. (\ref{eq:cov_matrix_NE_Langevin1}) yields
\begin{eqnarray}
\frac{\dd \Sigma}{\dd t}&=&A\Sigma+\Sigma A^\TT\nonumber \\ &+&\frac{\varepsilon_{\vartheta}^{\alpha}}{\alpha}\ee^{A t}\left(\ee^{-At}\Sigma\ee^{-A^\TT t}\right)^{1-\alpha}\left[\ee^{-At}(2D)^{\frac{1}{\alpha}}\ee^{-A^\TT t}\right]^\alpha\ee^{A^\TT t},
\end{eqnarray}
where we have assumed a time independent diffusion matrix. Since we are assuming $A$ stable, the equilibrium covariance matrix is then obtained through the following matrix equation:
\begin{eqnarray}
A\Sigma_{\rm eq}+\Sigma_{\rm eq} A^\TT+\frac{2\varepsilon_\vartheta^{\alpha}(\Sigma_{\rm eq})^{1-\alpha}D}{\alpha}=0.
\end{eqnarray}
For $\alpha=1$, we recover the continuous in time Lyapunov equation \cite{Simoncini16}. $\theta$-Gaussian processes whose expectation value and covariance matrix are obtained from Eqs. (\ref{eq:qexpectation_value_NE_Langevin}) and (\ref{eq:cov_matrix_NE_Langevin1}), with both $D$ and $b$ being time independent and $A$ stable, are then NE versions of multivariate Ohrnstein-Uhlenbeck processes. A particularly relevant case occurs when $A$ is proportional to the identity. Denoting $A=-\gamma \mathbb{I}_d$, Eq. (\ref{eq:cov_matrix_NE_Langevin1}) equals
\begin{eqnarray}
\Sigma(t)=\varepsilon_{\vartheta}\bigg[\frac{D}{\alpha\gamma}\big(1-\ee^{-2\alpha \gamma t}\big)\bigg]^{\frac{1}{\alpha}},
\end{eqnarray}
which asymptotically converges to $\Sigma_{\rm eq}=\varepsilon_{\vartheta}(D/\alpha\gamma)^{\frac{1}{\alpha}}$ 

We now proceed by showing how these family of generalized ILEs are related with nonlinear Fokker-Planck equations (FPEs). First, consider $\eta=1$. Under this restriction, the PDF $p$ is a time dependent $q$-Gaussian distribution (\ref{eq:q_Gaussian}) with zero expectation value and covariance matrix satisfying Eq. (\ref{eq:Sigma_anomalous_diff}) with $\alpha=1+d\lambda(1-q)$, if we assume a point-source initial condition $p(x,0)=\delta(x)$. The PDFs satisfy the following power-law-type nonlinear FPE: 
\begin{eqnarray}\label{eq:nonlinear_FPE1}
\frac{\partial p}{\partial t}+\diver\left[pF-(v_{2-q}[p])^{1-2\lambda}D\nabla p^{2-q}\right]=0,
\end{eqnarray}
if we define $\varepsilon_{\vartheta}$ to be
\begin{eqnarray}\label{eq:epsilon}
\varepsilon_{\vartheta}=\frac{1}{2_{qd}\pi_{qd}}\left[2(2-q)\alpha\hspace{0.05cm}\pi_{qd}\right]^\alpha\hspace{0.5cm} (\eta=1).
\end{eqnarray}
In the particular case where $\lambda=1/2$, Eq. (\ref{eq:nonlinear_FPE1}) reduces to the well-known porous media equation \cite{Vazquez, Aronson}
\begin{eqnarray}\label{eq:PME}
\frac{\partial p}{\partial t}=\diver\big(pF-D\nabla p^{2-q}\big).
\end{eqnarray}
The exponent characterizing the anomalous diffusion is
\begin{eqnarray}
\alpha^{-1}=\frac{2}{2+d(1-q)}\hspace{0.5cm} (\eta=2\lambda=1),
\end{eqnarray}
which was firstly derived by Tsallis-Bukman \cite{TB96} through the maximization of Tsallis non-additive entropy \cite{Tsallis88} and experimentally verified within great precision (2\% of error) in a granular media diffusion experiment \cite{CRSA15}. The corresponding stochastic process, in turn, was firstly studied in Ref. \cite{Santos21}. If $\lambda\neq 1$, Eq. (\ref{eq:nonlinear_FPE1}) is related with the Sharma-Mittal entropy \cite{FD00,KL12a,KL12b,KD21}, a two parameter non-additive generalization of BG entropy that has Tsallis entropy as a particular case \cite{Frank}.

\begin{figure*}
    \centering
    \includegraphics[scale=0.325]{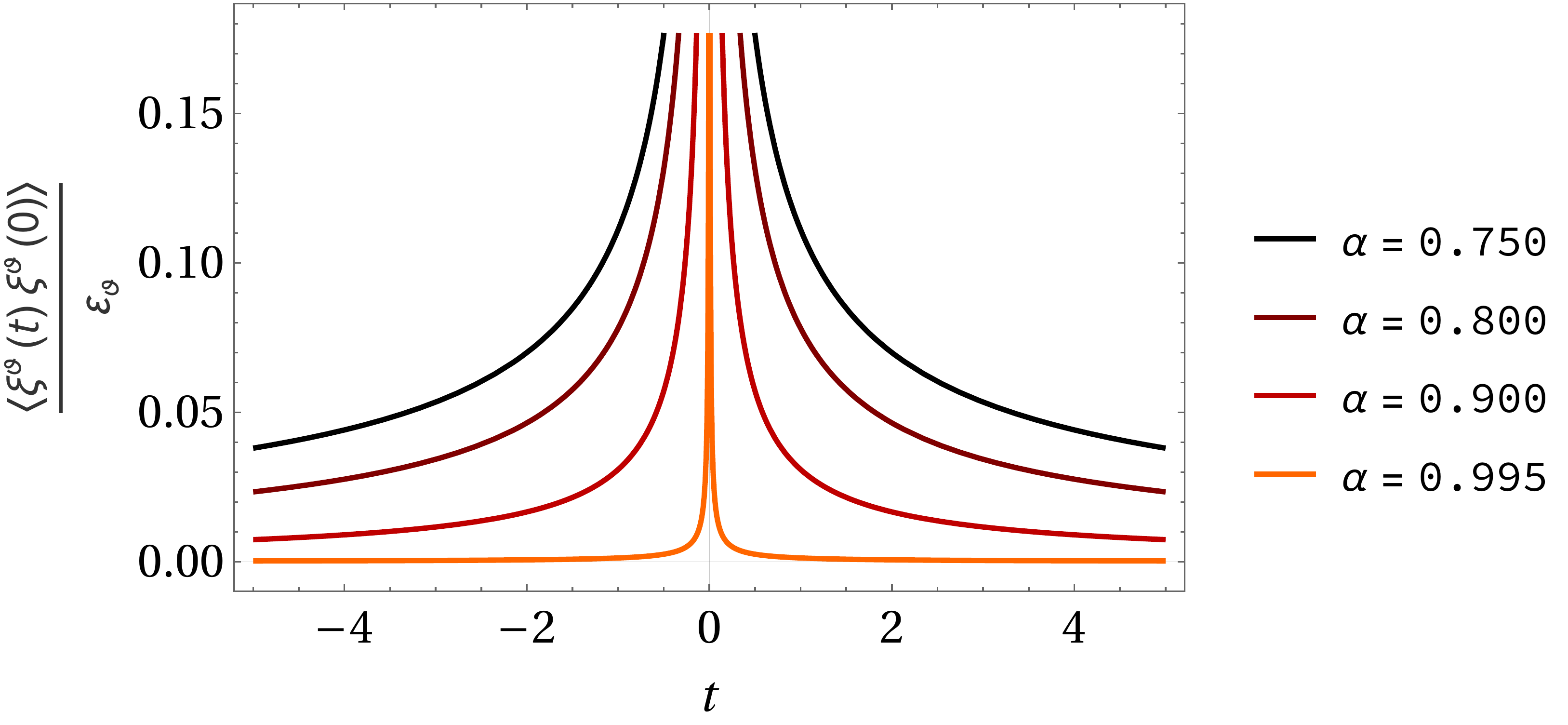}
    \caption{Time correlations for different values of $\alpha$ in the super-diffusive regime for $d=1$ [Eq. (\ref{eq:correlations})].}
    \label{fig:corr}
\end{figure*}

Now consider $q=2\lambda=1$ but $\eta\neq 1$. For simplicity, let us assume that $D$ is proportional to the identity and $\varepsilon_\vartheta=\eta^2$. The corresponding generalized ILE yields a stochastic process $X_t$ whose PDF is
\begin{eqnarray}
p(x,t)=\frac{2\eta}{\Gamma(d/2\eta)(4\eta^2 Dt)^{d/2\eta}}\exp\bigg(-\frac{\|x\|^{2\eta}}{4\eta^2 Dt}\bigg),
\end{eqnarray}
where $D$ is now a positive number. This PDF is the solution of the O’Shaughnessy-Procaccia equation \cite{SP85a,SP85b}
\begin{eqnarray}\label{eq:SPEq}
\frac{\partial p}{\partial t}=\frac{D}{r^{d-1}}\frac{\partial}{\partial r}\left(r^{d+1-2\eta}\frac{\partial p}{\partial r}\right).
\end{eqnarray}
If $d$ is interpreted as a positive real number, it plays the role of a fractal dimension embedded in some $N$-dimensional Euclidean space. The anomalous diffusion exponent in this case is $\alpha^{-1}=\eta^{-1}$. If we consider $q>1$, we can mix the correlated anomalous diffusion regime described by nonlinear FPEs (\ref{eq:nonlinear_FPE1}) and O’Shaughnessy-Procaccia diffusion in fractals by introducing a power-law nonlinearity in Eq. (\ref{eq:SPEq}). In this case, the solution is a $\theta$-Gaussian distribution and the anomalous diffusion exponent $\alpha^{-1}$ equals
\begin{eqnarray}\label{eq:alpha2}
\alpha^{-1}=\frac{2}{2\eta+d(1-q)}.
\end{eqnarray}
This problem has been studied in detail in Refs. \cite{MMPL01,PMML02} from the point of view of FPEs, where $\theta$-Gaussian distributions arise from a Barenblatt-Pattle-type ansatz (scale invariant solutions) \cite{Barenblatt,Pattle}. Our formalism then provides a microscopic model for this scenario and extends it with the introduction of the parameter $\lambda$, which may control the memory of the process.

In the traditional Langevin prescription, Eq. (\ref{eq:ito_langevin}) is often written as
\begin{eqnarray}\label{eq:Langevin}
\frac{\dd x}{\dd t}=F(x,t)+\sqrt{2D}\xi(t), 
\end{eqnarray}
where $\xi(t)=(\xi_1(t),\dots,\xi_d(t))^\TT$ is the $d$-dimensional Gaussian white noise, heuristically defined through $\xi(t)\dd t=\dd W_t$. The memoryless action of the stochastic force is read by the fact that $\xi$ exhibits delta correlation, i.e. $\langle\xi_a(t)\xi_b(s)\rangle=d\delta_{ab}\delta(t-s)$. It is possible to define the NE analog of $\xi$, $\xi^\vartheta(t)=(\xi_1^\vartheta(t),\dots,\xi_d^\vartheta(t))^\TT$, as $\xi^\vartheta(t)\dd t=\dd W_t^\vartheta$. Now, we have memory effects characterized by the fact that the noise $\xi^{\vartheta}$ is not $\delta$-correlated but instead
\begin{eqnarray}\label{eq:correlations}
\langle \xi_a^\vartheta(t)\xi_b^\vartheta(s)\rangle= d\delta_{ab}\frac{\varepsilon_{\vartheta}(1-\alpha)}{4\alpha^2}|t-s|^{-\frac{2\alpha-1}{\alpha}},
\end{eqnarray}
as a consequence of Eq. (\ref{eq:cova_maintext}). The ``tail" presented in these correlators quantifies the temporal correlations of the noise (see Fig. \ref{fig:corr}). Furthermore, the signal of $\langle \xi_a^\vartheta(t)\xi_b^\vartheta(s)\rangle$ explains super and sub diffusive regimes. Indeed, notice that if $\langle \xi_a^\vartheta(t)\xi_b^\vartheta(s)\rangle$ is positive(negative) then the action of the stochastic force at an instant $t+\Delta t$, $\Delta t>0$, is positively(negatively) correlated with the action at a previous instant $t$, tending to amplify(reduce) it. In this case, we would expect a super(sub) diffusive regime, a fact that is consistent with our discussion: the correlator $\langle \xi_a^\vartheta(t)\xi_b^\vartheta(s)\rangle$ is positive(negative) if, and only if, $\alpha^{-1}>1$($\alpha^{-1}<1$). Furthermore, Eqs. (\ref{eq:correlations}) and (\ref{eq:anomalous_exp}) show us that our formalism verifies the Hurst's law with Hurst exponent
\begin{eqnarray}\label{eq:Hurst}
H=\frac{1}{2\alpha},
\end{eqnarray}
as the fractional Brownian motion \cite{MvN68}. In terms of the $\theta$-Gaussian colored noise $\xi^\vartheta$, Eq. (\ref{eq:Langevin}) reads
\begin{eqnarray}\label{eq:qLangevin}
\frac{\dd x}{\dd t}=F(x,t)+(2D)^{\frac{1}{2\alpha}}\xi^\vartheta(t).
\end{eqnarray}
The above Langevin equation for $x(t)$ considers an additive colored noise whose solutions can be extracted through the formalism presented in this section. If $\eta=2\lambda=1$, the same dynamical behavior is obtained from the following Langevin-type equation (cf. Ref. \cite{Borland98})
\begin{eqnarray}\label{eq:StochasticBorland}
\frac{\dd x}{\dd t}=F(x,t)+\sqrt{2D}[p(x,t)]^{\frac{1-q}{2}}\xi(t),
\end{eqnarray}
where $p$ is obtained through Eq. (\ref{eq:PME}). In the above equation, we have a multiplicative noise for the stochastic dynamics of $x(t)$, which is described in terms of the Gaussian white noise $\eta$. In a sense, the coupling between microscopic and macroscopic dynamics, i.e. the fact that Eq. (\ref{eq:StochasticBorland}) explicitly depends on $p$, arises because we are trying to describe a non-Markovian and non-Gaussian system in terms of Gaussian white noise (or equivalently, the Wiener process).

\begin{figure*}
    \centering
    \includegraphics[scale=0.25]{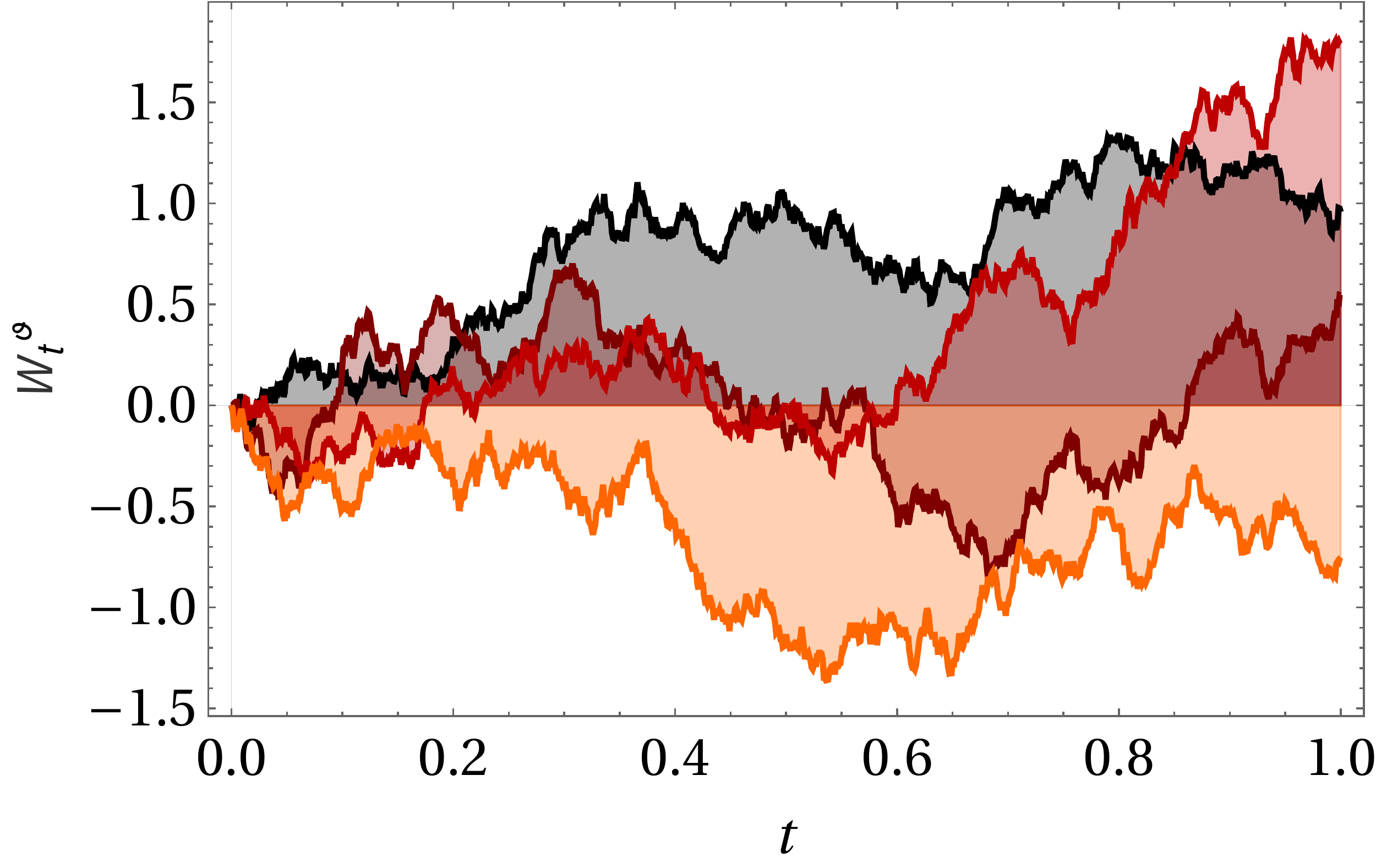}
    \includegraphics[scale=0.25]{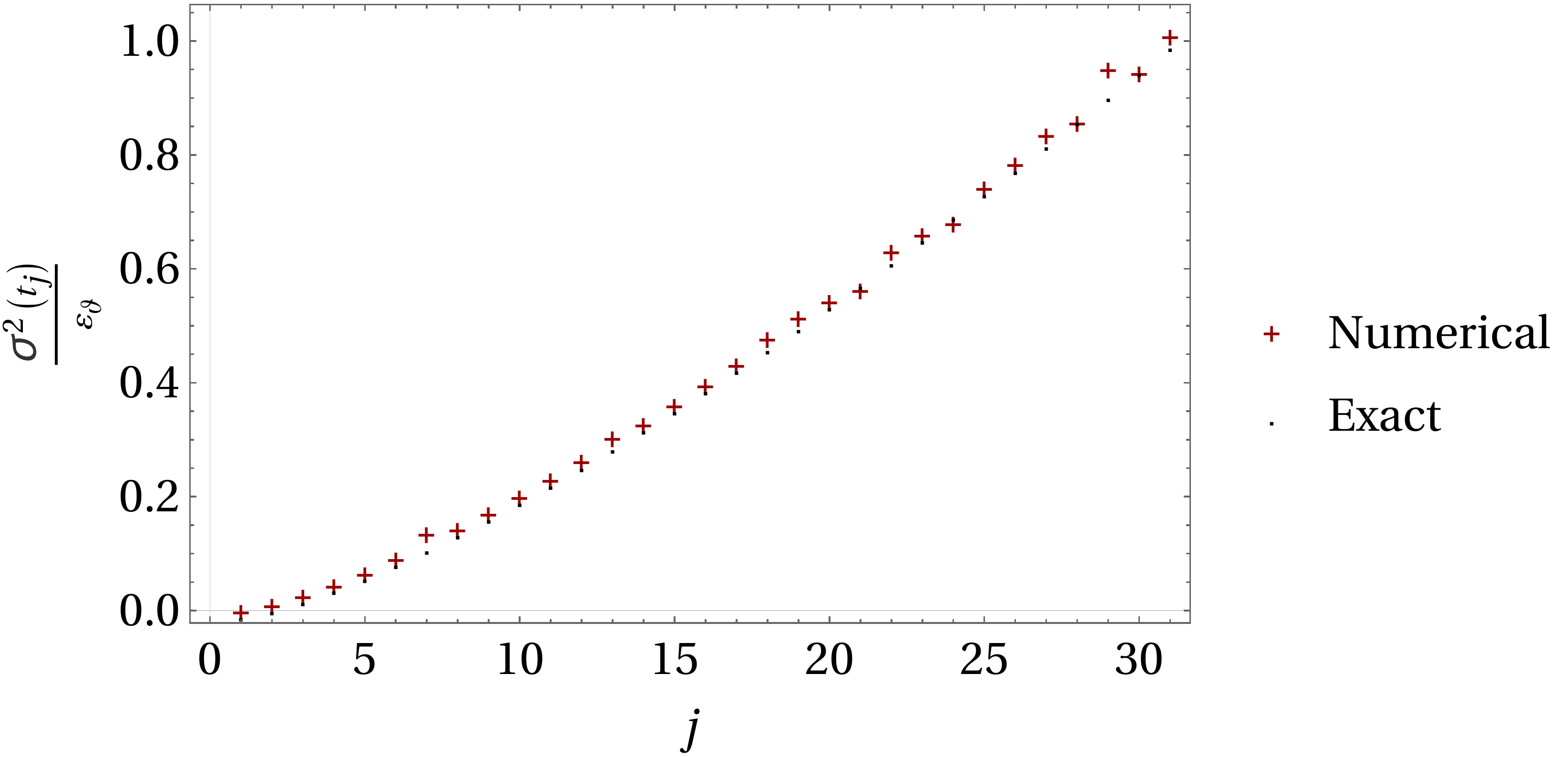}
    \caption{Stochastic trajectories of $W_t^{\vartheta}$ (left) and numerical estimation of variance (right) with $q=1.5$, $\eta=1$, $d=1$ and $\lambda=1/2$.}
    \label{fig:strajectories}
\end{figure*}

We now proceed by showing how the stochastic trajectories of $W_t^{\vartheta}$ can be generated. For this notice that, as a consequence of It\^{o}'s lemma \cite{Privault13} together with the definition of $W_t^{\vartheta}$, we have the following equality in distribution:
\begin{eqnarray}
\dd W_t^{\vartheta}=\sqrt{\dd t} L(W_t^{\vartheta},t)\mathfrak{g}_t,
\end{eqnarray}
where, for each $t$, $\mathfrak{g}_t\sim N(0,1)$ and 
\begin{align}
L(x,t)=&\|x\|^{2(\eta-1)}\Big\{v_{2-q}\Big[G_{\theta}\big(-;0,\varepsilon_{\vartheta} t^{\frac{1}{\alpha}}\big)\Big]\Big\}^{2\lambda-1}\nonumber \\ &\times G_{\theta}\big(x;0,\varepsilon_{\vartheta} t^{\frac{1}{\alpha}}\big).
\end{align}
We can thus generate the stochastic trajectories of $W_t^\vartheta$, e.g. from $t=0$ to $t=1$, by generating $n$ independent Gaussian random variables $\mathfrak{g}_j$ (e.g., through the Box-Muller method) in such a way that
\begin{eqnarray}
W_{j+1}^\vartheta=\frac{L(W_{j}^\vartheta,j/n)}{\sqrt{n}}\mathfrak{g}_j
\end{eqnarray}
together with $W_0^{\vartheta}=0$ define a sequence of random variables that approximates a realization of $W_t^{\vartheta}$ in the interval $[0,1]$.

\section{Discussion}\label{sec:4}

\noindent The goal of this paper was to study nonextensive generalizations of It\^{o}-Langevin dynamics. We believe this is of value for the following reasons. First, nonextensive statistical mechanics is a fundamental theory underlying many complex system phenomena in physics and beyond \cite{Tsallis09}. This includes those phenomena associated with $q$-Gaussian distributions and nonlinear Fokker-Planck equations, as discussed in the throughout this paper. In our contribution we have developed a consistent generalization of the standard It\^{o}-Langevin formalism. This allows us to extract analytical solutions in a simple manner and other properties such as time-correlations [Eqs. (\ref{eq:cova_maintext}) and (\ref{eq:correlations})], Hurst exponent [Eq. (\ref{eq:Hurst})], and self-similarity laws [Eq. (\ref{eq:self_similarity})] of stochastic processes describing correlated anomalous diffusion in fractals.

From a more theoretical or mathematical point of view, nonextensive statistical mechanics also provides significant insights \cite{UTBook}. In particular, we highlight the role of generalized central limit theorems to explain why there are so many $q$-Gaussian-like distributions in nature \cite{Tsallis09a}. Our contribution in this direction was to extend the results of Ref. \cite{UTS08}. This generalization was needed to properly justify the emergency of generalized Wiener processes. We believe that this is indeed a significant step towards a definite theory explaining the appearance of $q$-Gaussian distributions. Nevertheless, we have shown that strict $\vartheta$-independence is a sufficient condition for the emergency of $q$-Gaussian random variables. It is then natural to search for necessary conditions [at this point, possible extensions of the formalism for $(q,\alpha)$-stable distributions of Ref. \cite{UTGS10} would be warmly welcome]. Other equally interesting directions to investigate are the scale-invariant probabilistic models and large deviation theory (see, e.g., Refs. \cite{RST08,HTT09,STRT15,TTA21}), that have been shown promise in explaining the emergence of $q$-Gaussians, and their possible connection with the results developed here. 
\section*{Acknowledgments}

\noindent The author thanks Prof. Constantino Tsallis, Prof. Otfried Gühne, Prof. Stefan Nimmrichter, Prof. Gabriel Landi. Prof. Jorge Anderson Ramos, Prof. M\'{a}rcio Bortoloti and Carlos de Gois for their insightful comments and suggestions. This work was supported by the Conselho Nacional de Pesquisa Científica (CNPq) (Grants No. GM:155496/2019-0 and GD:141212/2021-7), the Deutsche Forschungsgemeinschaft (DFG, German Research Foundation, project numbers 447948357 and 440958198), the Sino-German Center for Research Promotion (Project M-0294) and the ERC (Consolidator Grant 683107/TempoQ).

\newpage
\textrm{ }
\newpage

\appendix
\begin{widetext}
\section{Technicalities of Section 2}\label{app:A} 

\subsection{Generalized Gaussian Distributions and Central Limit Theorems}\label{app:a1}

\noindent In section 2, we presented the $\theta$-Gaussian distributions and discussed how it can emerge as a limit of correlated random variables. In this supplemental material we shall discuss the technical part of this section. First, $\theta$-Gaussian distributions were defined as
\begin{eqnarray}\label{eq:theta_Gaussian_app1}
G_{\theta}(x;\mu,\Sigma)=\frac{\exp_q\left\{-\frac{1}{2_{\theta}}[(x-\mu)^\TT\Sigma^{-1}(x-\mu)]^\eta\right\}}{\sqrt{(2_\theta\hspace{0.05cm} \pi_\theta)^d\det\Sigma}},
\end{eqnarray}
where $\theta=(q,\eta,d)\in\mathbb{R}^2\times\mathbb{N}$ and $\exp_q(x)=[\max\{1+(1-q)x,0\}]^{\frac{1}{1-q}}$. The normalization constants $2_\theta$ and $\pi_\theta$ can be defined through
\begin{eqnarray}
\pi_\theta=\left(\int_{\mathbb{R}^d} \ee_q^{-\|x\|^{2\eta}}\dd x\right)^{\frac{2}{d}}\hspace{1cm}\textrm{and}
\hspace{1cm}
2_\theta=d(\pi_\theta)^{\frac{d}{2}}\left(\int_{\mathbb{R}^d}\|x\|^2\hspace{0.05cm}\ee_q^{-\|x\|^{2\eta}}\dd x\right)^{-1}.
\end{eqnarray}
It is easy to verify that $2_\theta\to 2$ and $\pi_\theta\to \pi$ when $q=\eta\to 1$, as stated in the main text. Given $\vartheta=(\theta,\lambda)$, $\lambda\in\mathbb{R}$, we defined the $\vartheta$-characteristic function of a random variable $X$ with probability density function (PDF) $p$ as
\begin{eqnarray}
\mathrm{F}_{X}^{\hspace{0.05cm}\vartheta}(k)=\int_{\supp p}p(x)\exp_q\left\{\ii {k}^{\TT} {x}[p(x)]^{\frac{1-\alpha}{d}}\right\}\dd x,
\end{eqnarray}
where $\alpha=\eta+d(1-q)$. In the main text we mention two properties of this integral transform. First, the scaling property:
\begin{eqnarray}\label{eq:scaling}
\mathrm{F}_{rX}^{\hspace{0.05cm}\vartheta}(k)=\mathrm{F}_{X}^{\hspace{0.05cm}\vartheta}(r^{\alpha}k),
\end{eqnarray}
$r>0$. This is a direct consequence of the fact that the PDF of $rX$ is $\tilde{p}(x)=r^{-d}p(x/r)$, where $p$ is the PDF of $X$. The second property is related with $\theta$-Gaussian random variables. Let $X\sim N_\theta(0,\Sigma)$. The $\vartheta$-characteristic function of $X$ is
\begin{eqnarray}\label{eq:FTeq1}
\mathrm{F}_{X}^{\hspace{0.05cm}\vartheta}(k)&=& \int_{\mathbb{R}^d} G_\theta(x;0,\Sigma)\exp_q\left\{\ii {k}^{\TT} {x}[G_q({x};{0},{\Sigma})]^{\frac{1-\alpha}{d}}\right\}\dd{x}\nonumber \\ &=&\int_{\mathbb{R}^d}\frac{\ee_q^{-\|{u}\|^{2\eta}}}{(\pi_\theta)^{\frac{d}{2}}}\left\{1-\ii(q-1)\left(a\ee_q^{-\|u\|^{2\eta}}\right)^{\frac{1-\alpha}{d}} \tilde{{k}}^{\TT}{u}\right\}^{\frac{1}{1-q}}\dd {u} \nonumber \\ &=&\mathrm{Exp}_{\theta}\left(-\frac{2_\theta}{4(2_\theta\pi_\theta)^{1-\alpha}}\frac{{k}^{\TT}{\Sigma}{k}}{(\det{\Sigma})^{\frac{1-\alpha}{d}}}\right),
\end{eqnarray}
where
\begin{eqnarray}\label{eq:FTeq2}
\mathrm{Exp}_{\vartheta}(x)=\frac{1}{(\pi_\theta)^{\frac{d}{2}}}\int_{\mathbb{R}^d}\ee_q^{-\|{u}\|^{2\eta}}\left[1-2\sqrt{x}(q-1)(\ee_q^{-\|u\|^{2\eta}})^{\frac{1-\alpha}{d}}v^{\TT}{u}\right]^{\frac{1}{1-q}}\dd {u},
\end{eqnarray}
$x\in\mathbb{R}$, and ${v}$ an arbitrary norm-1 vector in $\mathbb{R}^d$. In Eq. (\ref{eq:FTeq1}) we have used the fact $\mathrm{F}^{\hspace{0.05cm}\vartheta}_{{X}}$ depends only on the norm square of $\ii a^{\lambda(q-1)} \tilde{{k}}$. This is a direct consequence of the fact that the odd escort moments of the $\theta$-Gaussian distribution $\propto\ee_q^{-\|x\|^{2\eta}}$ are zero. From Eq. (\ref{eq:FTeq2}) we directly conclude that $\mathrm{Exp}_{\vartheta}(x)$ is non-negative and $\mathrm{Exp}_{\vartheta}(0)=1$, where the former fact is a consequence of the power-law-type decay of $\ee_q^{-\|x\|^{2\eta}}$ for $q>1$ and the latter follows from the normalization of the $\theta$-Gaussian.

In the particular case where $X$ is a $q$-Gaussian random variable and $\lambda=1$, Eq. (\ref{eq:FTeq1}) equals
\begin{eqnarray}\label{eq:UTS_qFT}
\mathrm{F}_{{X}}^{\hspace{0.05cm}q}({k})&=& \frac{1}{(\pi_{qd})^{\frac{d}{2}}}\int_{\mathbb{R}^d}\ee_q^{-\|{u}\|^2}\left\{1-\frac{\ii(q-1)a^{q-1} \tilde{{k}}^{\TT}{u}}{1+(q-1)\|{u}\|^2}\right\}^{\frac{1}{1-q}}\dd {u} \nonumber \\ &=&\left[1-(1-q)\frac{2_{qd}}{4(2_{qd}\pi_{qd})^{d(q-1)}}\frac{{k}^{\TT}{\Sigma}{k}}{(\det{\Sigma})^{q-1}}\right]^{\frac{2+d(1-q)}{2(1-q)}} \nonumber\\ &=& \exp_{\tilde{q}_d}\left\{-\frac{2_{qd}[2+d(1-q)]}{8(2_{qd}\pi_{qd})^{d(q-1)}}\frac{{k}^{\TT}{\Sigma}{k}}{(\det{\Sigma})^{q-1}}\right\},
\end{eqnarray}
where $\tilde{q}_d=[2+(d+2)(1-q)]/[2+d(1-q)]$ and $\mathrm{F}^{\hspace{0.05cm}q}$ denotes the UTS' $q$-characteristic function. This is the multivariate version of the Lemma 2.5 of Ref. \cite{UTS08} (cf. Ref. \cite{UT07}). In this particular case, the $q$-FT of a $q$-Gaussian is a $\tilde{q}_d$-Gaussian (up to normalization) if $q\geq 1$. In the cases where $q<1$, it was demonstrated by Umarov-Queir\'{o}s \cite{UQ10} that the $q$-FT of a $q$-Gaussian in general does not correspond to another $q$-Gaussian [even with a different value of $q$ as in Eq. (\ref{eq:UTS_qFT})]. In our discussion, the undesirable behavior presented by $\theta$-FTs when $q<1$ is that the corresponding $\vartheta$-Exponential can assume negative values.

\begin{figure}
    \centering
    \includegraphics[angle=0,scale=0.4]{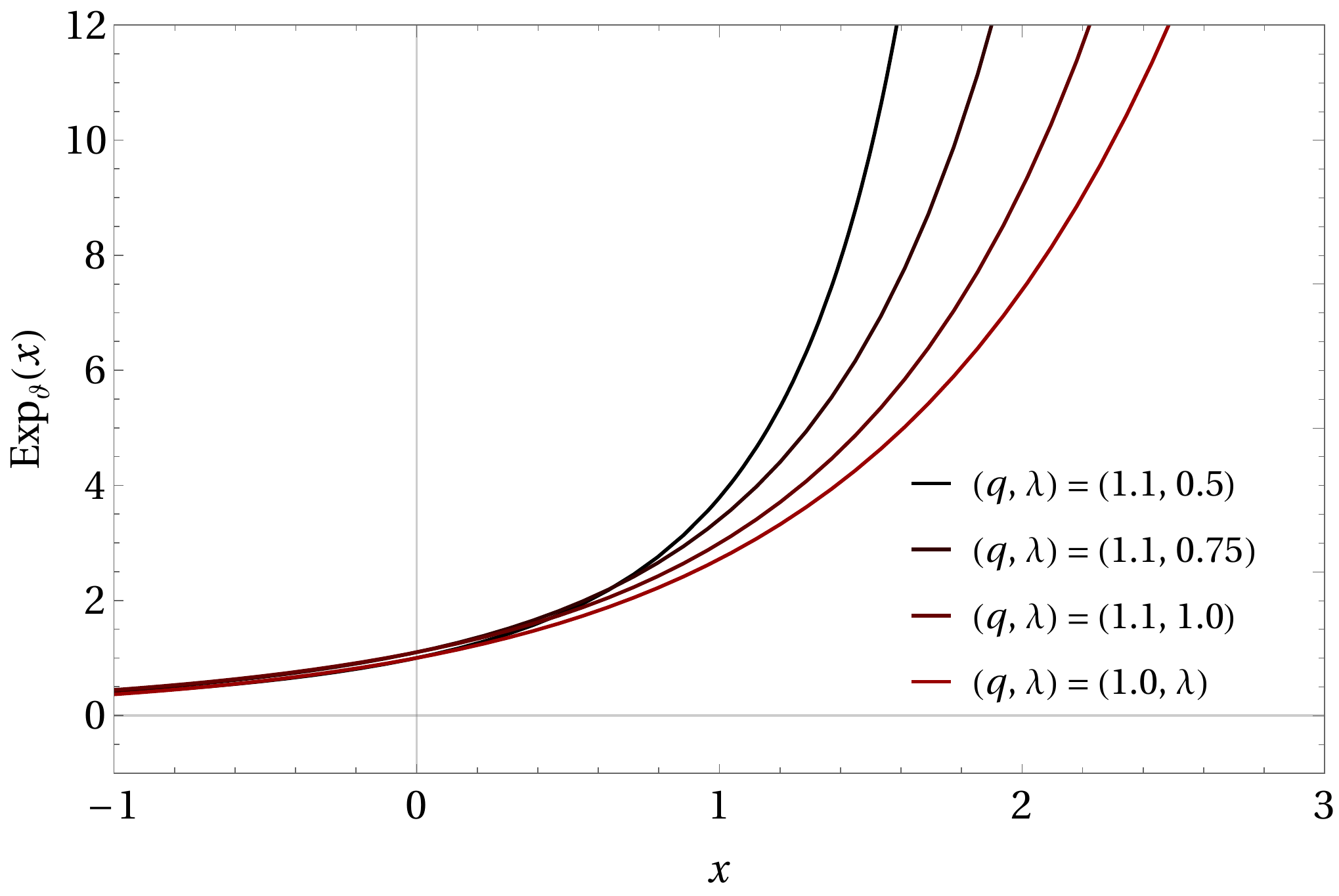}
    \caption{Generalized $\vartheta$-exponential with $d=\eta=1$ [Eq. (\ref{eq:FTeq2})].}
    \label{fig:theta_Exp}
\end{figure}

We have introduced in Eq. (\ref{eq:FTeq2}) a new deformed exponential function, the $\vartheta$-exponential.  This is a three parameter generalization of the exponential function $\ee^x$ (which is recovered when $q=\eta\to 1$). In general it is not easy to find closed expressions for $\vartheta$-exponential. As illustration, we compute the $\vartheta$-exponentials corresponding to some values of $\vartheta$ (see Fig. \ref{fig:theta_Exp}). First, if $\eta=\lambda=1$ we have
\begin{eqnarray}
\mathrm{Exp}_{(q,\hspace{0.05cm}d,\hspace{0.05cm}1,\hspace{0.05cm}1)}(x)=\exp_{\tilde{q}_d}\left[\frac{2+d(1-q)}{2}x\right],
\end{eqnarray}
$\tilde{q}_d=[2+(d+2)(1-q)]/[2+d(1-q)]$ [cf. Eq. (\ref{eq:UTS_qFT})]. Similarly, if $\eta=d=2\lambda=1$
\begin{align}
\mathrm{Exp}_{(q,\hspace{0.05cm}1,\hspace{0.05cm}1,\hspace{0.05cm}1/2)}(x)&={ }_{2}F_1\left(\frac{1}{2(q-1)},\frac{q}{2(q-1)},\frac{1}{q-1},4(q-1)x\right),
\end{align}
where ${ }_2 F_1$ denotes the Gaussian hypergeometric function. Finally, if $q=d=1$ we have
\begin{eqnarray}
\mathrm{Exp}_{(1,\hspace{0.05cm}1,\hspace{0.05cm}\eta,\hspace{0.05cm}\lambda)}^{\hspace{0.05cm}q}(x)=\frac{1}{\Gamma\left(\frac{2\eta+1}{2\eta}\right)}\int_{0}^\infty \ee^{-|u|^{2\eta}} \exp\left(2 u \sqrt{x}\hspace{0.05cm}\ee^{-(1-\eta)|u|^{2\eta}}\right)\dd u.
\end{eqnarray}
As the exponential, for all $\lambda>0$ and $\eta=1$, $\mathrm{Exp}_\vartheta(x)>0$, $\mathrm{Exp}_\vartheta(0)=1$ and $x\geq y$ implies $\mathrm{Exp}_\vartheta(x)\geq \mathrm{Exp}_\vartheta(y)$. For $\lambda\leq 0$ or $\eta\neq 1$, in turn, $\mathrm{Exp}_\vartheta(x)>0$ only if $x\leq 0$ and $\mathrm{Exp}_\vartheta(x)=0$ if $x>0$. We define the $\vartheta$-Logarithm as the inverse of $\mathrm{Exp}_{\vartheta}$ in the appropriated domain. The usual (natural) logarithm is recovered when $q\to 1$, while for $\lambda=1$ we have, in terms of the $\tilde{q}_d$-logarithm, $\mathrm{Log}_\vartheta (x)=2[2+d(1-q)]^{-1}\log_{\tilde{q}_d} (x)$, where $\log_q$ is the $q$-logarithm of NE statistical mechanics \cite{Tsallis09}
\begin{eqnarray}
\log_q u=\frac{u^{q-1}-1}{q-1}.
\end{eqnarray}

From the above results, we would conclude that $\vartheta$-Fourier transform (FT) are natural NE extensions of the usual FT. The proposed generalization, however, does not have all the good properties of its extensive counterpart as already noted for the UTS' $q$-FT. One the most mentioned problems is its (lack of) invertibility, since it is not invertible in the full space of PDFs for $q>1$. In fact, as demonstrated by Hilhorst \cite{Hilhorst10}, the following PDFs
\begin{eqnarray}
H_{r;\hspace{0.05cm}q}(x)=\begin{cases}\frac{\big(|x|^{\frac{q-2}{q-1}}-r\big)^{\frac{1}{q-2}}}{\sqrt{2_q\pi_{q}}|x|^{\frac{1}{q-1}}}\exp_q\left[-\frac{1}{2_q}\left(|x|^{\frac{q-2}{q-1}}-r\right)^{\frac{2(q-1)}{q-2}}\right]\hspace{0.1cm}&\textrm{if } |x|^{\frac{q-2}{q-1}}>r \\ 0 \hspace{0.25cm}& \textrm{otherwise}\end{cases},
\end{eqnarray}
$1<q<2$ and $r\in\mathbb{R}$, share the same UTS' $q$-FT of $G_q(x;0,1)$, $q>1$. This occurs because the $\vartheta$-FT associates a family of functions to each real function in $L_1$, instead only one as the usual FT. By using the same arguments of Refs. \cite{JT11,ST17}, it is possible to demonstrate that 
\begin{eqnarray}\label{eq:invertion}
f({x})=\left(\frac{1}{c_{\vartheta}}\int_{\mathbb{R}^d}\int_{\mathbb{R}^d}f({y})\ee_q^{\ii {k}^{\TT}({y}-{x})[f({y})]^{\frac{1-\alpha}{d}}}\dd {y}\dd {k}\right)^{\frac{1}{\alpha}},
\end{eqnarray}
for all ${x}$ in the interior of the support of $f$, $1\leq q<(1+d)/d$, $c_{\vartheta}>0$.
Eq. (\ref{eq:invertion}) shows that in order to recover the original function $f$ we need to know its $\vartheta$-FT and all possible translations of it. The only exception occurs when $q=\eta\to1$, since in this case we can factorize the exponential in the argument of the integral obtaining the inverse of the usual FT.

With these definitions on hand, we are now ready to generalize CLT.

\begin{thm}
Let $X_n$ be a sequence of $\vartheta$-independent and identically distributed random variables. Suppose also
\begin{eqnarray}\label{eq:condition1}
\int_{\supp p} \|{x}\|^2[p({x})]^{s_2}\dd {x}<\infty,
\end{eqnarray}
where $p$ denotes the PDF of ${X}_n$. Then the sequence $\tilde{X}_n$,
\begin{eqnarray}\label{eq:condition2}
\tilde{X}_n=(n z)^{-\frac{1}{2\alpha}}\left(S_n-n E_{s_1}[X_n]\right),
\end{eqnarray}
$z>0$, $S_n=X_1+\dots +X_n$, $\vartheta$-converges to a $\theta$-Gaussian random variable ${X}$ with zero expectation value and covariance matrix ${\Sigma}$ satisfying
\begin{align}\label{eq:covmatrix}
\frac{\Sigma}{(\det\Sigma)^{\frac{1-\alpha}{d}}}=\frac{2_{\theta'}v_{s_2}[p]}{2_{\theta}(2_{\theta}\pi_{\theta})^{\alpha-1}}\left(\frac{\pi_{\theta} s_2}{\pi_{\theta'}}\right)^{\frac{d}{2}}E_{s_2}[X_1X_1^\TT],
\end{align}
where $\mathbb{I}_d$ denotes the $d\times d$ identity matrix and $s_n=1+n(1-\alpha)/d$.
\end{thm}
\begin{proof}
From condition (\ref{eq:condition2}), we can (without loss of generality) assume $E_{s_1}[X_i]=0$ for all $i\geq 1$, since $E_{s_1}[X-E_{s_1}[X]]=0$ for any random variable $X$. The $\vartheta$-characteristic function of $\tilde{X}_n=(nz)^{-\frac{1}{2\alpha}}(X_1+\dots+X_n)$ then reads
\begin{eqnarray}\label{eq:thm3_eq1}
\mathrm{F}^{\hspace{0.05cm}\vartheta}_{\tilde{{X}_n}}({k})&=&\mathrm{F}^{\hspace{0.05cm}\vartheta}_{{X}_1}({k}/\sqrt{nz})\otimes_\vartheta \dots \otimes_{\vartheta} \mathrm{F}^{\hspace{0.05cm}\vartheta}_{{X}_n}({k}/\sqrt{nz}),
\end{eqnarray}
where we have made use of the scaling property of Eq. (\ref{eq:scaling}) together with the fact that the sequence is supposed to be $\vartheta$-independent. Since the sequence is also supposed to identically distributed, taking the $\vartheta$-Log in both sides of Eq. (\ref{eq:thm3_eq1}) we obtain
\begin{eqnarray}
\mathrm{Log}_{\vartheta}\big[\mathrm{F}^{\hspace{0.05cm}\vartheta}_{\tilde{{X}}_n}({k})\big]=n\mathrm{Log}_{\vartheta}\big[\mathrm{F}^{\hspace{0.05cm}\vartheta}_{{{X}}_j}\big({k}/\sqrt{nz}\big)\big].
\end{eqnarray}
With some manipulation, the $\vartheta$-characteristic function can be expanded as
\begin{align}\label{eq:A10}
\mathrm{F}^{\hspace{0.05cm}\vartheta}_{{{X}}_j}\big({k}/\sqrt{nz}\big)=1-\frac{qv_{s_2}[p]}{2n z}{k}^{\TT}E_{s_2}[XX^\TT]{k}+\mathcal{O}\big(n^{-3/2}\big),
\end{align}
where we have made use of the fact $E_{s_1}[{X}_j]={0}$.
The $\theta$-Exponential, in turn, can be expanded as
\begin{align}
\mathrm{Exp}_{\vartheta}(y)&=\int_{\mathbb{R}^d} \frac{\ee_q^{-\|{x}\|^{2\eta}}}{(\pi_{\theta})^{\frac{d}{2}}}\exp_q\left[2\sqrt{y}({x}^{\TT}{v})\left(\ee_q^{-\|{x}\|^{2\eta}}\right)^{\frac{1-\alpha}{d}}\right]\dd {x} \nonumber \\ &=1+(2qy){v}^{\TT}\left(\int_{\mathbb{R}^d}\frac{\left(\ee_q^{-\|{x}\|^{2\eta}}\right)^{s_2}}{(\pi_{\theta})^{\frac{d}{2}}}{x}{x}^\TT \dd {x}\right){v}+\mathcal{O}(y^2) \nonumber\\ &=1+(2qy)\int_{\mathbb{R}^d}\frac{\left(\ee_q^{-\|{x}\|^{2\eta}}\right)^{s_2}}{(\pi_{\theta})^{\frac{d}{2}}}x_j^2 \dd {x}+\mathcal{O}\big(y^2\big) \nonumber \\ &=1+\frac{2q}{2_{\theta'}}\left(\frac{\pi_{\theta'}}{\pi_{\theta}s_2}\right)^{\frac{d}{2}} y+\mathcal{O}\big(y^2\big),
\end{align}
where ${v}$ is an arbitrary norm-1 vector in $\mathbb{R}^d$ and $\theta'=(q',\eta,d)$, $q'=[s_2-(1-q)]/s_2$. Comparing the above equation with Eq. (\ref{eq:A10}), in the limit $n\to \infty$ we obtain the following expression for the $\theta$-characteristic function of ${X}$:
\begin{align}
\mathrm{F}^{\hspace{0.05cm}\vartheta}_{{X}}({k})=\mathrm{Exp}_{\vartheta}\left\{-\frac{2_{\theta'}v_{s_2}[p]}{4z}\left(\frac{\pi_{\theta}s_2}{\pi_{\theta'}}\right)^{\frac{d}{2}}k^\TT E_{s_2}\big[X_1 X_1^\TT\big]k\right\}.
\end{align}
Comparing the above equation with Eq. (\ref{eq:FTeq1}), we therefore identify ${X}$ as a $\theta$-Gaussian random variable with zero expectation value and covariance matrix given by
\begin{eqnarray}
\frac{\Sigma}{(\det\Sigma)^{\frac{1-\alpha}{d}}}=\frac{2_{\theta'}v_{s_2}[p]}{2_{\theta}(2_{\theta}\pi_{\theta})^{\alpha-1}}\left(\frac{\pi_{\theta} s_2}{\pi_{\theta'}}\right)^{\frac{d}{2}}E_{s_2}[X_1X_1^\TT]
\end{eqnarray}

\end{proof}

\begin{rem}
It is worthy to mention that the lack of invertibility of $q$-generalized Fourier transforms and its impact on the corresponding CLTs have been widely discussed in literature for the UTS formalism. We highlight, in particular, the discussion of Ref. \cite{UT16}, where the authors presented various arguments stating the uniqueness of limit distributions for the UTS' CLTs. Fortunately, the results and conclusions presented in such an article can be easily adapted to the formalism presented here, since the $\vartheta$-FT has almost the same form of the $q$-FT.
\end{rem}

The usual CLT is recovered when $q\to 1$ whereas for $\lambda=1$ we have a multivariate version of UTS' CLT \cite{UTS08}, with the limit covariance matrix being (cf. Ref. \cite{UT07})
\begin{eqnarray}
\frac{{\Sigma}}{(\det{\Sigma})^{q-1}}=\frac{4q(2_{qd}\pi_{qd})^{d(q-1)}}{2_{qd}[2+d(1-q)]z}\int_{\supp p}[p({x})]^{2q-1}{x}{x}^\TT \dd {x},
\end{eqnarray}
and a $q$-Gaussian distribution.

\subsection{Generalized Wiener Process}\label{app:a2}

\noindent Let us now demonstrate the existence of a well-defined stochastic process ${Q}_t^\vartheta$, constructed as a limit of $\vartheta$-independent random variables, for $\eta=1$, $1\leq q<\min\{(\lambda d+1)/\lambda d,(d+4)/(d+2)\}$ and $\lambda>0$, satisfying:
\begin{enumerate}[label=(\arabic*)]
\item\label{item:i} ${Q}_0^\vartheta={0}$ almost surely;
\item\label{item:ii} The paths $t\mapsto{Q}_{t}^\vartheta$ are continuous with probability one;
\item\label{item:iii} ${Q}_{t}^\vartheta-{Q}_{s}^\vartheta\sim N_q\big[0,(t-s)^{\frac{1}{\alpha}}\mathbb{I}_d\big]$, $0\leq s\leq t$;
\item\label{item:iv} ${Q}_{t}^\vartheta-{Q}_s^\vartheta$, $0\leq s\leq t$, is $\vartheta$-independent of ${Q}_s^\vartheta$;
\item\label{item:v} Given $t,s\geq 0$, the covariance between the components of ${Q}_t^\vartheta$ and ${Q}_s^\vartheta$ is \begin{eqnarray}\label{eq:covariance}
\cov[Q_{t;\hspace{0.05cm}\mu}^{\vartheta},Q_{s;\hspace{0.05cm}\nu}^{\vartheta}]&:= \frac{\delta_{\mu\nu}}{2}\left(t^{\frac{1}{\alpha}}+s^{\frac{1}{\alpha}}-|t-s|^{\frac{1}{\alpha}}\right).
\end{eqnarray}
\end{enumerate}
First we will discuss the existence of such a stochastic process. For this, let
\begin{eqnarray}
p({x}_1,t_1;\dots;{x}_n,t_n)\equiv G_{\hat{q}_{nd}}\big({x}_1,\dots,{x}_n;{\Lambda}(t_1,\dots,t_n)\big),
\end{eqnarray}
define a joint PDF, $x_i\in\mathbb{R}^d$, where $0\leq t_1<t_2<\dots<t_n$; ${\Lambda}(t_1,\dots,t_n)$ is a $nd\times nd$ matrix defined as
\begin{eqnarray}
{\Lambda}(t_1,\dots,t_n)\equiv \left(\begin{matrix} {\lambda}(t_1; \hspace{0.01cm}t_1) & {\lambda}(t_1; \hspace{0.01cm}t_2)& \dots & {\lambda}(t_1; \hspace{0.01cm}t_n) \\  {\lambda}(t_2; \hspace{0.01cm}t_1) & {\lambda}(t_2; \hspace{0.01cm}t_2)& \dots & {\lambda}(t_2; \hspace{0.01cm}t_n) \\ \vdots & \vdots & \ddots & \vdots \\ {\lambda}(t_n; \hspace{0.01cm}t_1) & {\lambda}(t_n; \hspace{0.01cm}t_2)& \dots & {\lambda}(t_n; \hspace{0.01cm}t_n) \end{matrix}\right),
\end{eqnarray}
$({\lambda}(t_\mu,t_\nu))_{ab}=\cov[Q_{t_\mu;\hspace{0.05cm}a}^\vartheta,Q_{t_\nu;\hspace{0.05cm}b}^\vartheta]$ given by Eq. (\ref{eq:covariance}), and
\begin{eqnarray}
\hat{q}_{nd}=\frac{2q-d(n-1)(1-q)}{2-d(n-1)(1-q)}.
\end{eqnarray}
A simple calculation show that $p(x,t)=G_q(x;\hspace{0.05cm}0,t^{\frac{1}{\alpha}}\mathbb{I}_d)$ and \begin{subeqnarray}
\int_{\mathbb{R}^d} p({x}_1,t_1)\dd {x}_1&=&1, \\
p({x}_1,t_1;\dots;{x}_n,t_n) &\geq& 0, \\
\int_{\mathbb{R}^d} p({x}_1,t_1;\dots;{x}_n,t_n)\dd x_n&=&p({x}_1,t_1;\dots;{x}_{n-1},t_{n-1}), \\ p({x}_{\pi(1)},t_{\pi(1)};\dots;{x}_{\pi(n)},t_{\pi(n)})&=&p({x}_1,t_1;\dots;{x}_n,t_n),
\end{subeqnarray}
where $\pi$ is an element of the permutation group of $\{1,\dots,n\}$. These conditions allow us to apply the Kolmogorov consistency theorem \cite{Durrett10} for the induced probability measure, guarantying thus the existence of a (probably not unique) stochastic process $Q^\vartheta_t\sim N_q(0,t^{\frac{1}{\alpha}}\mathbb{I}_d)$ satisfying Eq. (\ref{eq:covariance}).

Consider now a sequence of $\vartheta$-independent random variables ${\xi}_j$ in $\mathbb{R}^d$ such that ${\xi}_j\sim N_q({0},\mathbb{I}_d)$, and let
\begin{eqnarray}
{X}_{t;\hspace{0.05cm}n}=n^{-\frac{1}{2\alpha}}\sum_{j=1}^{\lfloor nt\rfloor} {\xi}_j.
\end{eqnarray}
From generalized CLT, the sequence $({X}_{t;\hspace{0.05cm}n}:n\in\mathbb{N})$ $\vartheta$-converges to a random variable ${Q}^\vartheta_t$ with zero expectation value and covariance matrix satisfying
\begin{eqnarray}
\frac{{\Sigma}}{(\det {\Sigma})^{\lambda(q-1)}}=t \hspace{0.05cm}\mathbb{I}_d.
\end{eqnarray}
In the above equation we have made use of the fact that
\begin{align}
\frac{{2}_{q'd}({2}_{qd}\pi_{qd})^{\lambda d(q-1)}}{{2}_{qd}} \left\{\frac{\pi_{qd}[1+2\lambda(q-1)]^{\frac{2+d}{2}}}{\pi_{q'd}}\right\}^{\frac{d}{2}}\int_{\mathbb{R}^d} [G_q({x};{0},\mathbb{I}_d)]^{1+2\lambda(q-1)}{x}{x}^\TT \dd {x}=\mathbb{I}_d,
\end{align}
Therefore ${Q}_t^\vartheta\sim N_q\big({0},t^{\frac{1}{\alpha}}\mathbb{I}_d\big)$. In the same way, the sequence defined by ${X}_{t;\hspace{0.05cm} n}-{X}_{s;\hspace{0.05cm} n}$, $0\leq s\leq t$, $\vartheta$-converges to ${Q}_t^\vartheta-{Q}_s^\vartheta$. Moreover, since
\begin{eqnarray}
{X}_{n;\hspace{0.05cm} t}-{X}_{n;\hspace{0.05cm} s}&=&n^{-\frac{1}{2\alpha}}\sum_{i=\lfloor n s\rfloor}^{\lfloor n t\rfloor} {\xi}_k,
\end{eqnarray}
then ${Q}_t^\vartheta-{Q}_s^\vartheta$ is equal in distribution to ${Q}_{t-s}^\vartheta$. Therefore, ${Q}_{t}^\vartheta-{Q}_{s}^\vartheta\sim N_q\big[{0},(t-s)^{\frac{1}{\alpha}}\mathbb{I}_d\big]$. This, together with the following relation
\begin{eqnarray}
\cov[A,B]=\frac{1}{2}\left(\var[A]+\var[B]-\var[B-A]\right),
\end{eqnarray}
imply that the covariance between the components of ${Q}_{t}^\vartheta=\big(Q_{t;\hspace{0.05cm}1}^\vartheta,\dots,Q_{t;\hspace{0.05cm}d}^\vartheta\big)^\TT$ at different times equals Eq. (\ref{eq:covariance}).

We now have on hand a well-defined stochastic process ${Q}_t^\vartheta$ obeying conditions \ref{item:iii} and \ref{item:v}. Condition \ref{item:i} is also fulfilled since ${Q}_t^\vartheta\sim N_q\big({0},t^{\frac{1}{\alpha}}\mathbb{I}_d\big)$ and $G_q({x};{0},\varphi\mathbb{I}_d)\to \delta({x})$ when $\varphi\to 0$. Continuity of the paths of ${Q}^\vartheta_t$ [condition \ref{item:ii}], in turn, follows from Kolmogorov-Chentsov continuity theorem \cite{Chentsov}. In fact, the process $Q_t^\vartheta$ satisfies
\begin{eqnarray}
E[\|{Q}_t^\vartheta-{Q}_s^\vartheta\|^2]=d|t-s|^{1+\frac{d(q-1)}{\alpha}},
\end{eqnarray}
in such a way that for all $q$ such that $1<q<\min\{(d\lambda+1)/d\lambda,(d+4)/(d+2)\}$, there exists a modification of ${Q}^\vartheta_t$, $\tilde{{Q}}_t^\vartheta$, that is continuous, i.e. a sample-continuous stochastic process $\tilde{{Q}}_t^\vartheta$ such that
\begin{eqnarray}
\mathrm{Pr}[{Q}_t^\vartheta=\tilde{{Q}}_t^\vartheta]=1,
\end{eqnarray}
for all $t\geq 0$. Furthermore, the paths of $\tilde{{Q}}^\vartheta_t$ are $\beta$-Hölder continuous for all $\beta$ real such that 
\begin{eqnarray}
0<\beta<\frac{d\lambda(q-1)}{2\alpha}.
\end{eqnarray}
Therefore the continuous modification of the process ${Q}_t^\vartheta$ (which we shall also denote by the same symbol) is a well-defined stochastic process fulfilling conditions \ref{item:i}, \ref{item:ii}, \ref{item:iii}, and \ref{item:v}. Finally, the $\vartheta$-independence between ${Q}_t^{\vartheta}-{Q}_s^{\vartheta}$ and ${Q}_s^{\vartheta}$, $s<t$, follows straightforwardly:
\begin{eqnarray}
\mathrm{F}^{\hspace{0.05cm}\vartheta}_{({Q}_{t}^\vartheta-{Q}_{s}^\vartheta)+{Q}_s^\vartheta}({k})&=&\mathrm{F}^{\hspace{0.05cm}\vartheta}_{{Q}_{t}^\vartheta}({k}) \nonumber\\ &=&\mathrm{Exp}_{\vartheta}\left\{-\frac{{2}_{qd}}{4({2}_{qd}\pi_{qd})^{d\lambda(q-1)}}[(t-s)+s]\|{k}\|^2\right\} \nonumber \\ &=& \mathrm{F}^{\hspace{0.05cm}\vartheta}_{{Q}_t^\vartheta-{Q}_s^\vartheta}({k})\otimes_{\vartheta}\mathrm{F}^{\hspace{0.05cm}\vartheta}_{{Q}^\vartheta_s}({k}).
\end{eqnarray}

\subsection{Stochastic Integral}\label{app:a3}

\noindent To finish, let us discuss the stochastic integral induced by $Q_t^{\vartheta}$. The stochastic integral of a function $f$ with respect to ${Q}_t^\vartheta$ is a stochastic process ${I}_t$,
\begin{eqnarray}
{I}_t=\int_0^t f\big({Q}_\tau^\vartheta,\tau\big)\dd {Q}_\tau^\vartheta,
\end{eqnarray}
whose PDF is obtained by
\begin{eqnarray}
\mathrm{F}^{\hspace{0.05cm}\vartheta}_{{I}_t}({k})=\lim_{n\to \infty}\bigotimes_{\vartheta; \hspace{0.05cm} [t_{j-1},t_j]\in\mathsf{p}_n} \mathrm{F}^{\hspace{0.05cm}\vartheta}_{f({Q}^{\vartheta}_{t_{j-1}},\hspace{0.05cm}t_{j-1})\Delta {Q}_{t_j}^\vartheta}({k}),
\end{eqnarray}
where $\mathsf{p}_n$ is a sequence of partitions of the interval $[0,t]$ with mesh going to zero, $\Delta {Q}^\vartheta_{t_j}\equiv {Q}^\vartheta_{t_j}-{Q}_{t_{j-1}}^\vartheta$, and the product in the right-hand-side is the $\vartheta$-product indexed by the elements of $\mathsf{p}_n$. As claimed in the main text, this stochastic integral satisfies a generalized It\^{o} formula:
\begin{eqnarray}\label{eq:itoformulaappendix}
I_t\sim N_\theta\left[{0},\hspace{0.1cm}\left(\int_0^t |f(\tau)|^{2\alpha}\dd \tau\right)^{\frac{1}{\alpha}}\mathbb{I}_d\right],
\end{eqnarray}
where $f(t)\in L^{2\alpha}([0,t])$, $t\geq 0$. The proof of this result is straightforward:
\begin{align}
\mathrm{F}^{\hspace{0.05cm}\vartheta}_{{I}_t}({k})&=\lim_{n\to \infty}\bigotimes_{\vartheta; \hspace{0.05cm} [t_{j-1},t_j]\in\mathsf{p}_n} \mathrm{F}^{\hspace{0.05cm}\vartheta}_{f(t_{j-1})\Delta {Q}_{t_j}^\vartheta}({k}) \nonumber \\ &=\lim_{n\to \infty}\bigotimes_{\vartheta; \hspace{0.05cm} [t_{j-1},t_j]\in\mathsf{p}_n} \mathrm{F}^{\hspace{0.05cm}\vartheta}_{\Delta {Q}_{t_j}^\vartheta}\big(|f(t_{j-1})|^{\alpha}{k}\big) \nonumber \\ &=\lim_{n\to \infty}\bigotimes_{\vartheta; \hspace{0.05cm} [t_{j-1},t_j]\in\mathsf{p}_n} \mathrm{Exp}_{\vartheta}\left(-\frac{{2}_{\theta}}{4({2}_{\theta}\pi_{\theta})^{1-\alpha}}|f(t_{j-1})|^{2\alpha}\Delta t_j\|{k}\|^2\right) \nonumber \\ &=\mathrm{Exp}_{\vartheta}\left(-\frac{{2}_{\theta}}{4({2}_{\theta}\pi_{\theta})^{1-\alpha}}\lim_{n\to \infty}\sum_{[t_{j-1},t_j]\in\mathsf{p}_n}|f(t_{j-1})|^{2\alpha}\Delta t_j\|{k}\|^2\right) \nonumber\\ &=\mathrm{Exp}_{\vartheta}\left(-\frac{{2}_{\theta}}{4({2}_{\theta}\pi_{\theta})^{1-\alpha}}\int_0^t|f(\tau)|^{2\alpha}\dd \tau\hspace{0.05cm}\|{k}\|^2\right)
,\end{align}
where Eq. (\ref{eq:itoformulaappendix}) follows from direct comparison with Eq. (\ref{eq:FTeq1}).

\end{widetext}

\end{document}